\documentclass[]{spie}  

\usepackage{amsmath,amsfonts,amssymb}
\usepackage{graphicx}
\usepackage[colorlinks=true, allcolors=blue]{hyperref}

\def\arcmin{\hbox{$^\prime$}}
\def\arcsec{\hbox{$^{\prime\prime}$}}
\newboolean{english}
\newboolean{french}
\newboolean{german}
\newboolean{spanish}
\newboolean{italian}
\relax

\title{Optical aspects of Near-Infrared Imager Spectrometer and Polarimeter instrument (NISP)}

\author[a,b]{Archita Rai}
\author[a]{Shashikiran Ganesh}
\author[c]{Sukamal Paul}
\author[a]{Prashanth K Kasarla}
\author[a]{\\Prachi V Prajapati}
\author[a]{Deekshya R Sarkar}
\author[a]{Alka Singh}
\author[a]{Pitambar S Patwal}
\author[a]{\\Hitesh K L Adalja}
\author[a]{Satya N Mathur}
\author[a]{Sachindra Naik}
\author[a]{Amish B Shah}
\author[a]{Kiran S Baliyan}
\affil[a]{Physical Research Laboratory, Ahmedabad, India}
\affil[b]{Indian Institute of Technology, Gandhinagar, India}
\affil[c]{Space Application Center, Ahmedabad, India}

\authorinfo{Further author information: (Send correspondence to Archita Rai)\\Archita Rai: E-mail: archita@prl.res.in}

\begin{document} 
\maketitle

\begin{abstract}
As a Near- IR instrument to PRL's upcoming 2.5~m telescope, NISP is designed indigeniously at PRL to serve as a multifaceted instrument. Optical, Mechanical and Electronics subsystems are being designed and developed in-house at PRL. It will consist of imaging, spectroscopy \& imaging-polarimetry mode in the wavelength bands Y, J, H, Ks i.e. 0.8 – 2.5 $\mu m$. 
The detector is an 2K x 2K H2RG (MCT) array detector from Teledyne, which will give a large FOV of 10$^\prime$ x 10$^\prime$ in the imaging mode. Spectroscopic modes with resolving power of R $\sim$ 3000, will be achieved using grisms. 
Spectroscopy will be available in single order \& a cross- dispersed mode shall be planned for simultaneous spectra. 
The instrument enables multi-wavelength imaging- polarimetry using Wedged-Double Wollaston (WeDoWo) prisms to get single shot Stokes parameters (I, Q, U) for linear polarisation simultaneously, thus increasing the efficiency of polarisation measurements and reducing observation time. 
\end{abstract}


\section{INTRODUCTION}
\label{sec:intro}  
Multi-mode instruments are very important for a big telescope facility which combines several scientific goals to be achieved using single instrument. The modes of imaging, spectroscopy \& polarimetry are crucial for majority of the phenomena observed in astronomical bodies, and hence form the basis of implementation for an astronomical instrument. Their are many multi mode instruments around the world and they facilitate in serving extensively towards studies ranging from stellar to galactic, galactic to extra galactic, exoplanets, solar system and a lot more. The instruments like MIMIR\cite{mimir}, FORS\cite{fors}, SIRPOL\cite{sirpol}, TRISPEC\cite{trispec} etc have revolutionized for majority of scientific goals. Their are many survey mode instruments which are giving interesting results and have pitched in to answer some of the mysterious phenomena of our existence and formation taking us back to the Big bang and present.\\\\
The Indian facilities cover a range of telescopes at several locations, which include 3.6 m telescope at Devasthal, Nainital of ARIES; upcoming 2.5 m telescope at Mt.Abu, Rajasthan of PRL; 2 m telescope at HCT, Hanle of IIA; 2 m telescope at Girawali, Pune of IUCAA; 1.2 m telescope at Mt.Abu, Rajasthan of PRL and many more. These facilities house some of the advanced and indigeniously designed and fabricated instruments in optical and infrared wavelength bands, i.e., NICS \cite{NICS}, TIRSPEC \cite{tirspec}, TIRCAM2 \cite{tircam2}, TANSPEC, AIMPOL etc. The Indian landscape is also vividly rich in dry locations which are best suited for operating in infrared wavelength, naming Mt.Abu, Rajasthan \& Hanle, Ladakh as two such locations at an altitude of 1680 m and 4500 m respectively.\\\\
The further sections discuss the optical design development of a near-infrared instrument. The scientific goals which form the motivation behind the instrument are listed in the second section. The third section talks about the full instrument overview with each subsection discussing the various modes and their analysis. The fourth section explains the science performance by the full optics within the instrument. The fifth section summarises the achievements of the optical design.

\section{Scientific goals}
Science requirements are the major drivers for building of any new instrument. Its main aim is to define the specifications required for the performance which will need to be achieved for any particular scientific work. Being a general purpose instrument, a range of science goals have been listed for NISP.  A few example programs are pointed out below :\\
\begin{itemize}
    \item \textbf{Be/X ray binaries \& Novae :} X-ray binaries \& novae are the class of astronomical bodies which are associated with energetic outburts which involve physical processes spanning full electromagnetic spectrum. During normal Type I outbursts, the pulse profiles of Be/X-ray binary pulsars are found to be complex in soft X-ray energy ranges. Optical/infrared observations of the companion Be-star during these Type I outbursts showed that the increase in the X-ray intensity of the pulsar is coupled with the decrease in the optical/infrared flux of the companion star.

    \item \textbf{Solar system studies :} Comets and asteroids are extensively studied with the spectroscopy technique from low resolution $\sim$ 1000 to high resolution of $\sim$ 60,000 . One of the important NIR spectroscopic results expected is in the study of comets. The water ice feature present at 1.4 \& 2.0 $\mu m$ is a good tracer for the presence of water in comets.  The oxygen line seen in the visible spectrum of comets can arise from either H$_2$O or CO$_2$ molecules. Hence, compiling the data with NIR spectrum will help in concluding the origin of oxygen line from the water feature in comets.
    
    \item \textbf{Milky Way studies :} Imaging polarimetry of the highly extincted regions of the Milky Way galaxy will help us measure the polarization of redder stars which were obscured by the dust at shorter, visible wavelengths. Near IR bands have less scattering and hence can penetrate deeper into the obscured regions. With the polarization measurements, we can deduce the magnetic field geometry of the star forming cloud, the magnetic field strength \& infer dust properties.  ``Hidden" AGN can be detected in external active galaxies using the technique of imaging polarimetry.
\end{itemize}
\section{Instrument overview}
The instrument assembly constitutes of collimator and camera optics sections alongwith the filter wheels placed near the pupil area to carry out functioning of different modes. The re-imaging optics helps in transferring the image of the primary mirror on the pupil plane, which is further imaged by the camera assembly onto the detector. The width of the pupil plane decides the size of the optical components which will be used for carrying out different operations. The detector is a 2K x 2K H2RG array detector by Teledyne. 
\begin{figure}[h!]
    \centering
    \fbox{\includegraphics[width=0.8\textwidth]{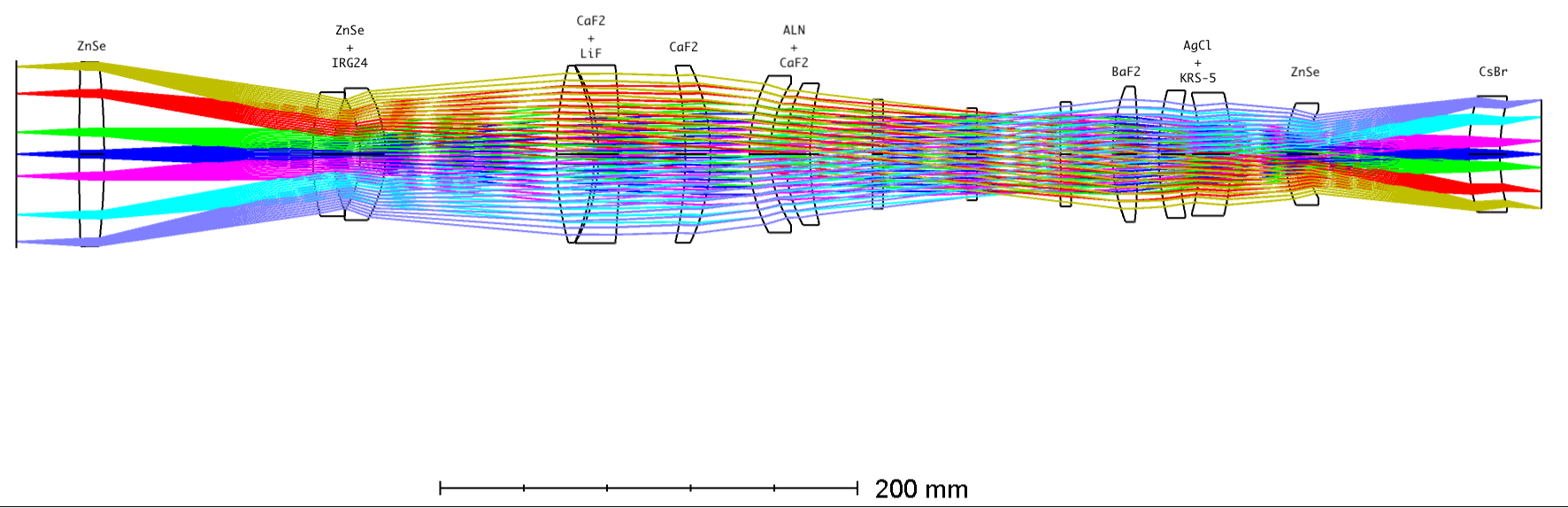}}
    \caption{Imaging optical design of NISP with the lens components labelled with the respective materials.}
    \label{imaging_layout}
\end{figure}
\subsection{Imaging optics}
The imaging optics of NISP consists of 8 big collimator lenses which collimate the $F/8$ beam from the telescope cassegrain focal plane. The field of view of the instrument in the imaging mode is 10$^{\prime}$ x 10$^{\prime}$. The collimated beam is of 38 mm and the different field rays cross at the location which is defined as the pupil plane. This is the best location to place grisms/wollastons. The camera optics succeeds the placement of filter wheels and is comparatively smaller is size. The $F/5$ optics of camera is designed to image the 10$^{\prime}$ x 10$^{\prime}$ field of view on the H2RG detector, thus implying a pixel scale for imaging of ${0.3}^{\prime \prime}$ per pixel. 
\begin{figure}[h!]
    \centering
    \fbox{\includegraphics[width=0.5\textwidth]{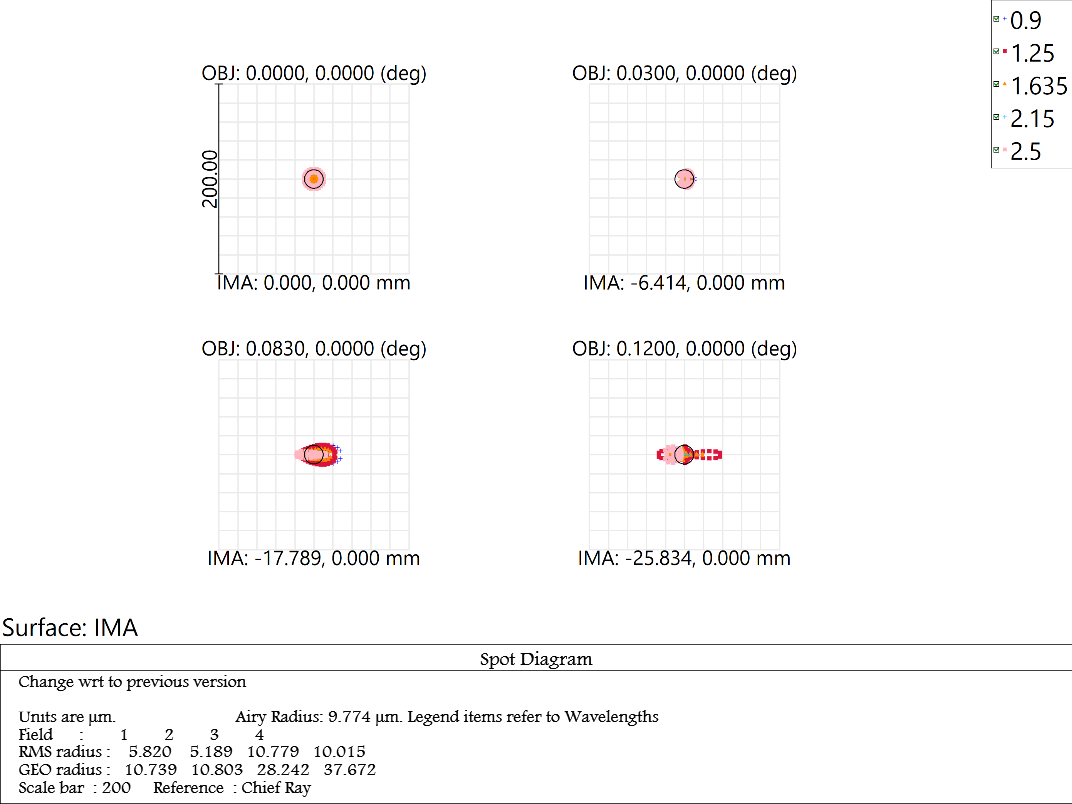}}
    \caption{Spot diagram for all the fields covering the $10^{\prime}$ x $10^{\prime}$ region on detector plane.}
    \label{spot_imaging}
\end{figure}
The layout in Fig.(\ref{imaging_layout}) shows the different fields used to map the full FOV on the cassegrain focal plane (on the left), and using the optics to image it on the comparatively smaller size detector chip. The design shows the materials which have been used for the optical design. All the lenses used in the optics are picked from infrared glasses catalogs available in Zemax. The total instrument optical assembly extends to a length of 740 $mm$. 
The design was brought to perfection of image by continuous attempt to achieve good spot size at all the defined field angles from the central chief ray to extreme. Fig.(\ref{spot_imaging}) shows the spot diagram plot from Zemax for the optical design. The value of Airy-disk radius is 9.86 $\mu m$ and the achieved RMS spot radius of 5.820 $\mu m$, 5.189 $\mu m$, 10.779 $\mu m$ \& 10.015 $\mu m$ at the respective fields are well within the required radius for a 18 $\mu m$ pixel. 
One of the important performance check for instrument designs quality is the Modulation transfer function. It defines the contrast which will be transferred from the total contribution of each optical element to the final detector. The value desirable for any instrument is the one closer to the diffraction limit performance. The plot in 
\begin{figure}[h!]
    \centering
    \fbox{\includegraphics[width=0.5\textwidth]{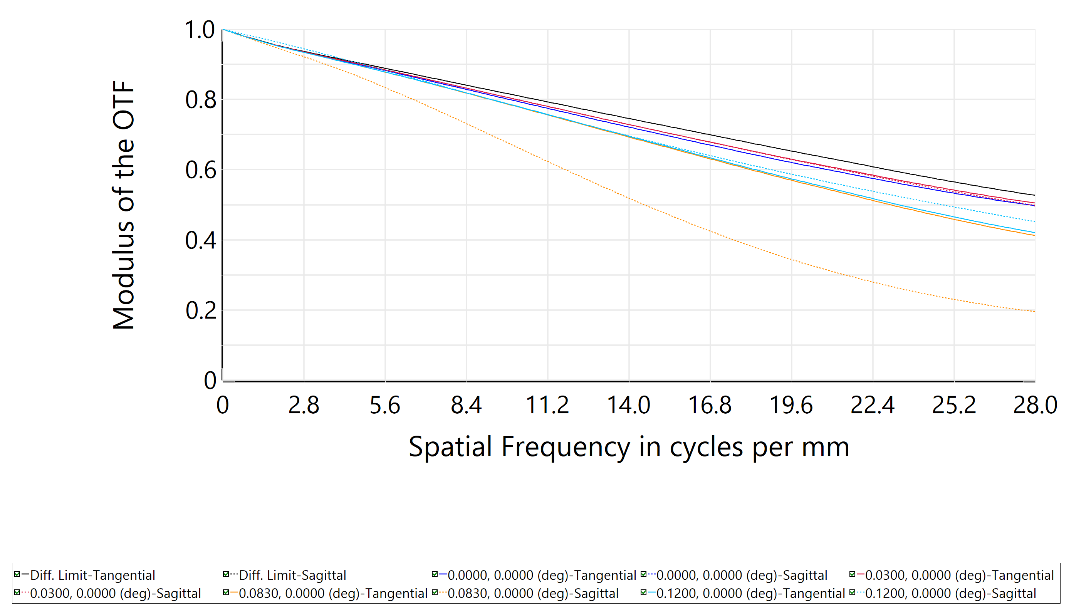}}
    \caption{MTF plot for imaging design of NISP}
    \label{MTF}
\end{figure}
Fig.(\ref{MTF}) shows the MTF over the spatial frequency decided by the line pairs/mm for NISP detector. The performance for almost all the fields is close to the bold black line marking the diffraction limit performance and is found to be above 40$\%$. 

\subsection{Spectroscopic optics}
\begin{figure}[h]
    \centering
    \fbox{\includegraphics[width=0.8\textwidth]{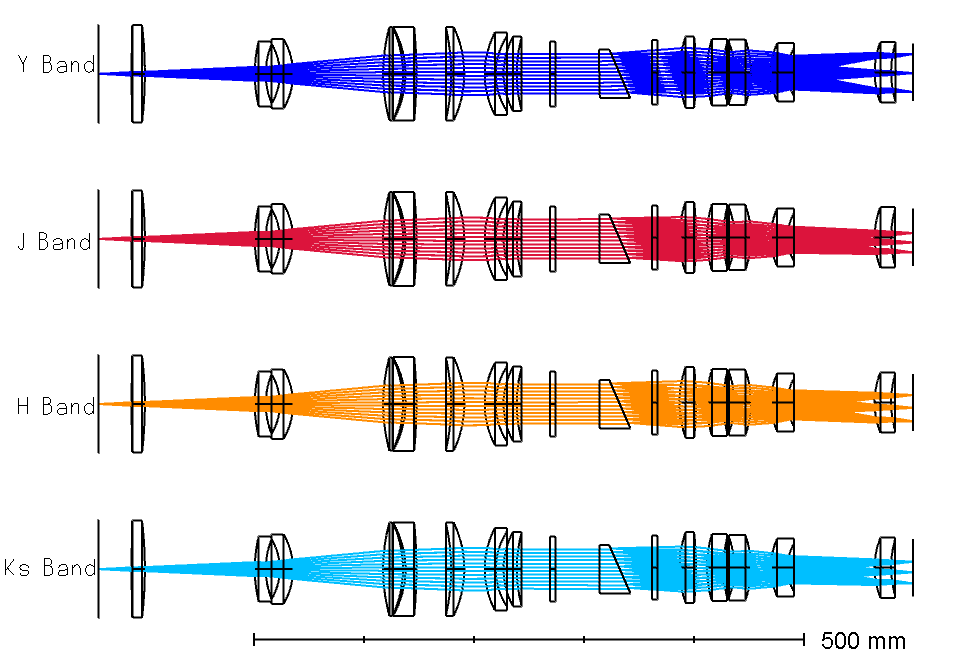}}
    \caption{Layout of the spectroscopy mode of NISP}
    \label{spect_layout}
\end{figure}

The technique of spectroscopy is inevitable in any general purpose astronomical instrument. All the telescope facilities are equipped with spectrographs of varying resolving powers. The main purpose of any spectrograph is to disperse the light falling on the slit and hence on the dispersing medium which can be a prism, grating or grism. The dispersed light is in the form of spectrum which is recorded on the detector and utilised to identify the compositions present in the source.\\
We use a grism for the spectroscopy purpose. There will be 4 grisms for each of the wavelength bands Y, J, H, K$_{s}$. The resolving power is 2150 in each of the 4 bands. A layout with the inclusion of grism is the optical path is shown in Fig.(\ref{spect_layout}). 
The Table \ref{grismcal} gives full description of the grism specifications used in the design. 
\begin{table}[h]
 \caption{The required parameters for an R = 2150, KRS5 material grisms in the Y, J, H, K$_s$ filters in the 1st order, with $F_{coll}$ = 339 mm, and $\Delta_x$ = 0.097 mm are tabulated.}
\centering
\begin{tabular}{|c|c|c|c|c|}
\hline
       Resolving Power  & Refractive Index & Wavelength  & Prism Angle  & Groove spacing  \\
       & & ($\mu m$) & ($^\circ$) & ($\mu m$ ) \\
       \hline \hline
      2150   &  2.4434 & 1.02 (Y) & 23.08 & 1.80\\ 
      2150 & 2.4209 & 1.25 (J) & 23.41 & 2.21\\ 
      2150 & 2.4031 & 1.635 (H) & 23.68 & 2.90 \\ 
      2150 & 2.3928 & 2.15 (K$_s$) & 23.83 & 3.82\\ 
\hline
\end{tabular}
\label{grismcal}
\end{table}

\subsection{Polarisation optics}
\begin{figure}[h]
    \centering
    \fbox{\includegraphics[width=0.86\textwidth]{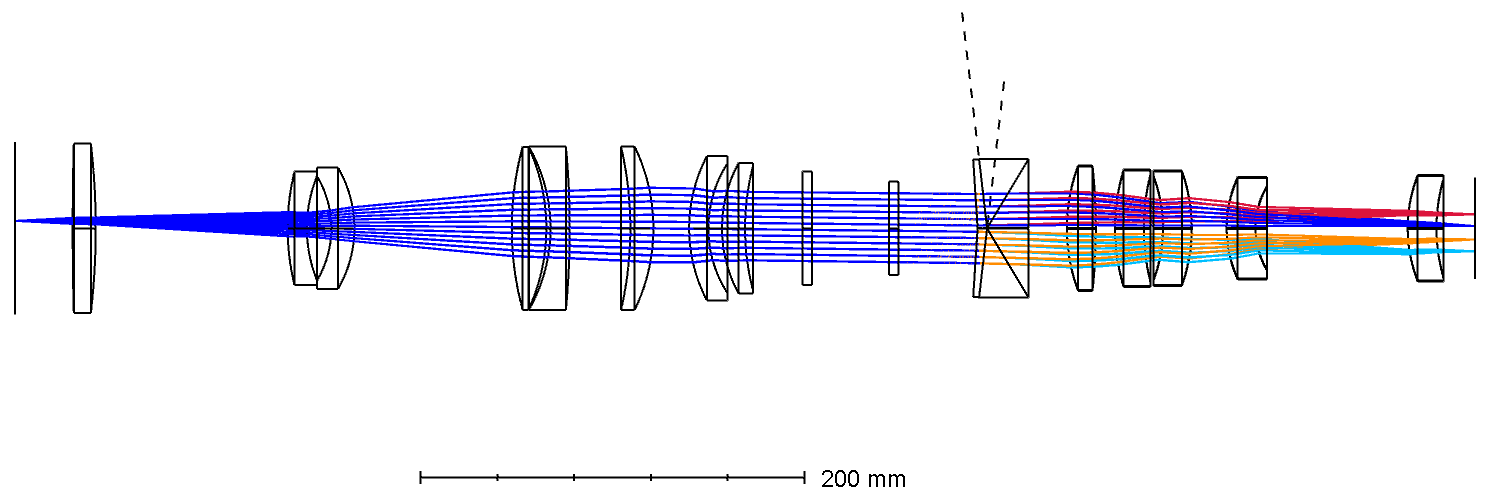}}
    \caption{Layout of the imaging polarimetry mode of NISP}
    \label{layout_pol}
\end{figure}

\begin{figure}[h]
    \centering
    \fbox{\includegraphics[width=0.7\textwidth]{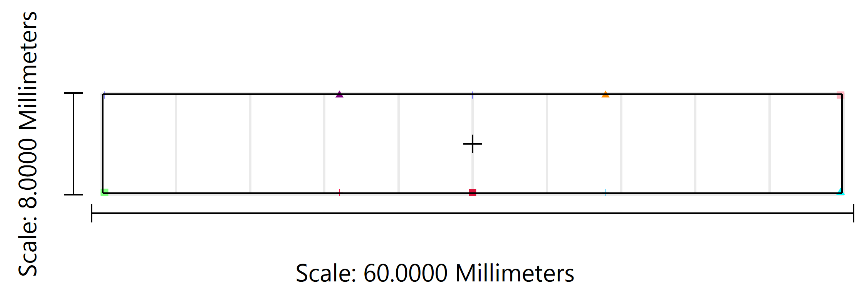}}
    \caption{Focal plane mask for imaging polarimetry mode of NISP}
    \label{focalmask}
\end{figure}

\begin{figure}[h]
    \centering
    \includegraphics[width=0.4\textwidth]{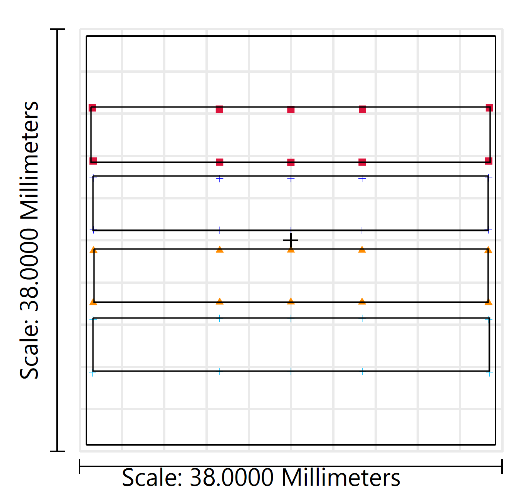}
    \caption{The polarisation images as horizontal strips of 10$^\prime$ x 1.4$^\prime$ FOV each at 0$^{\circ}$, 45$^{\circ}$, 90$^{\circ}$, 135$^{\circ}$ angles.}
    \label{polimage}
\end{figure}

The polarisation optics makes the use of Wollaston prism to utilise both the ordinary and extra-ordinary beams of light. The imaging polarimetry mode will be having double wollastons, i.e. WeDoWo (wedged double wollaston) to achieve four simultaneous images at different polarizing angles. The layout using the WeDoWo is shown in Fig.\ref{layout_pol}, and the 4 beams consisting of 2 ordinary \& 2 extra-ordinary are seen getting focused on detector plane. 
\begin{figure}[h]
    \centering
    \fbox{\includegraphics[width=0.35\textwidth]{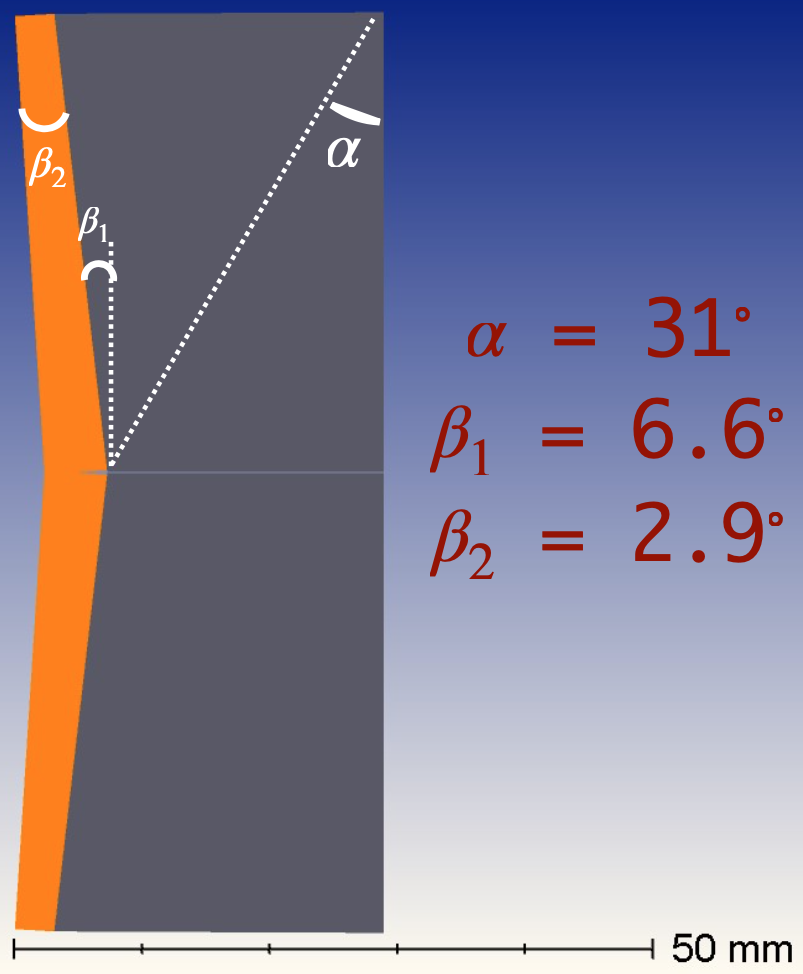}}
    \caption{WeDoWo designed using LiYF4 material for NISP with all the angles labelled.}
    \label{wedowo}
\end{figure}\\
The labelled design of the WeDoWo component is also shown in Fig.\ref{wedowo}. The angles calculation was done as explained in the Oliva et al (1997) \cite{oliva}. The material currently utilised is LiYF4 (YLF). As the YLF material has a low birefringence, the proper image separation requires a large Wollaston angle $\alpha$. Hence, the size of the WeDoWo component comes to $\sim$ 30 mm. We shall also use MgF2 material for the same purpose.\\
A focal plane mask is defined at the Cassegrain focus of the telescope. The dimensions of the mask are calculated from the idea of plate scale = 16.6${^\prime}{^\prime}$/mm. Hence, the focal plane mask defined to cover a rectangular FOV of 10$^\prime$ x 1.4$^\prime$, in linear scale is sized as 58.2 mm x 8 mm. The focal plane mask assembly in the instrument is shown in Figure \ref{focalmask}. Its placement is in the aperture wheel at the Cassegrain focus.
As can be seen from the footprint diagram in Figure \ref{polimage}, the full 36.8 mm square area is marked, and the horizontal strips define the images at the various angles of the Wollaston. The FOV in the imaging polarization mode is 10$^\prime$ x 1.4$^\prime$. 

\newpage
\section{Performance Analysis}
\begin{figure}[h]
    \centering
    \includegraphics[width=0.45\textwidth]{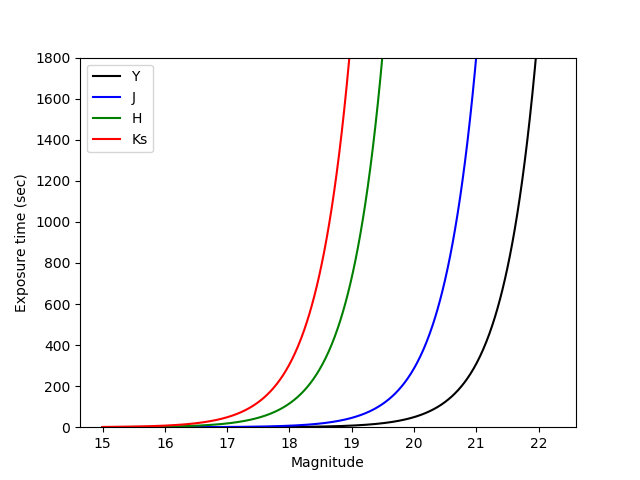}
   \caption{Exposure vs magnitude for SNR = 5 for NISP. }
    \label{exp_cal}
\end{figure}

\begin{figure}[h]
    \centering
    \fbox{\includegraphics[width = 0.8\textwidth]{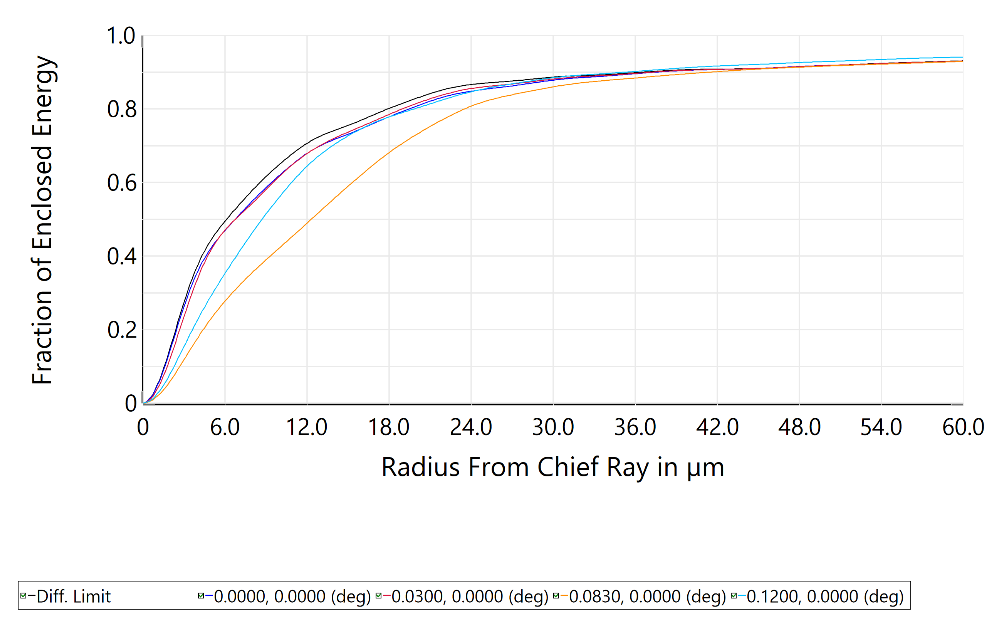}}
    \caption{Encircled energy plot for the imaging mode.}
    \label{ee_imaging}
\end{figure}

The optics performance was theoretically analysed for the efficiency achieved at the detector plane. Limiting magnitudes based on the current optical design and a few simplified assumptions for imaging mode are computed for a signal-to-noise ratio (SNR) of 5 and $1800$~sec of exposure. These values as shown in Fig.~\ref{exp_cal} are $22.5, 21.5, 19.7$, and $19.2$ for the broad-bands Y, J, H, and $K_s$ respectively. These NIR broad-band filters cover the wavelength range from $\rm 0.8 - 2.5~\mu m$.
The encircled energy is a fraction of the total integrated flux in the image contained within a given radius `r'. From the Fig. \ref{ee_imaging}, one can notice that the black color line is for the diffraction limited performance, and the other colors are meant for different field angles. It is evident that the distribution is mostly close to the diffraction limited performance. The size of one pixel being 18$\mu$m, 50\% energy should be within one pixel, and 80\% within 2 pixels. Since the plot~\ref{ee_imaging} has x-axis representing the radius from chief ray, 80$\%$ encircled energy is within 2 pixel (i.e. 18$\mu$m radius) for all the fields. 

\section{Summary}
The achieved design specifications are below:\\
\begin{table}[h]
\begin{tabular}{l l}
        Achieved Spot Size : & $< 54~ \mu$m \\
       Pupil diameter :  & 38 mm \\
       Field of View (FOV) : & $10^{\prime}$ x $10^{\prime}$ \\
       Resolving power : & 2150 \\
       Encircled emergy : & EE90 $\sim$ 2 pixels \\
    \end{tabular}
    \label{specs}
\end{table}\\
The summary from the optical design are :
\begin{enumerate}
\item The imaging design of NISP consists of a re-imager combination of collimator \& camera lenses to focus the incident light at various field positions covering the full FOV on the detector plane. An F/5 camera system is designed for the sampling of 1$\arcsec$ on 3 pixels of the detector. \\
\item The spectroscopic design uses a `grism' as the main component to disperse the different wavelength bands. Separate grisms for each of the 4 bands Y, J, H, K$_s$ had been designed using the KRS-5 material in the 1st order with a spectral resolving power of 2150. A slit of length 1.4$\arcmin$ \& width $\sim$ 1$\arcsec$ is placed in the aperture wheel at the Cassegrain focal plane.\\
\item The polarization design uses a WeDoWo component to ensure single shot simultaneous polarization measurements at 4 position angles. A focal plane mask of 10$\arcmin$ x 1.4$\arcmin$, has been designed to create 4 non-overlapping images corresponding to each of the angles.
\end{enumerate}

\bibliography{report} 
\bibliographystyle{spiebib} 

\end{document}